\documentstyle[manuscript,epsfig,osa]{revtex}

\begin{document}

\bibliographystyle{unsrt}

\title{\rmfamily Optics of a Faraday-active Mie Sphere\sffamily }

\author{D. Lacoste, B. A. Van Tiggelen \\  \, 
\em Laboratoire de Physique et Mod\'elisation des Milieux Condens\'es, Maison des Magist\`eres, \\
\em B.P. 166, 38042 Grenoble Cedex 09, France \\
G.L.J.A Rikken, and A. Sparenberg \\
\em Grenoble High Magnetic Field Laboratory, Max-Planck Institut \\
\em f\"ur Festk\"orperforschung/C.N.R.S, B.P. 166, 38042 Grenoble Cedex 9,France}
\maketitle

\begin{abstract}

We present an exact calculation for the scattering of light from a
single sphere made of a Faraday-active material, into first order of
the external magnetic field. When the size of the sphere is small compared
to the wavelength the known T-matrix for a magneto-active Rayleigh
scatterer is found. We address the issue whether or not 
there is a so called \em Photonic Hall Effect - \em a magneto-transverse
 anisotropy in light scattering - for one Mie scatterer.
In the limit of geometrical optics, we compare our results to the
Faraday effect in a Fabry-Perot etalon.

\end{abstract}

\bigskip


\section{Introduction}

Several reasons exist why one wishes to understand light scattering
from a dielectric sphere made of magneto-active material. Single scattering
is the building block for multiple scattering. Many experiments
 have been done with diffuse light in a magnetic field
 \cite{lenke,nature,anja}. Though 
qualitatively very useful, it turns out that a
theory for point-like scatterers in a magnetic field, as first developed
by MacKintosh and John \cite{sajeev} and later refined by Van Tiggelen,
 Maynard and Nieuwenhuizen \cite{bart}, does not always describe
observations quantitatively, for the evident reason that experiments
do not contain ``small'' scatterers. This paper addresses light scattering
from one sphere of any size in a homogeneous magnetic field.

The model of Rayleigh scatterers has been successfully used to describe
 specific properties of multiple light scattering in magnetic
fields such as coherent backscattering and the \em Photonic Hall Effect \em.
 In the following section we present the perturbation approach 
 on which our work is based, which will allow us in section 3 to compute the
  T-matrix for this problem. With this tool, we will be able to answer
   in section 4 the issue concerning the \em Photonic Hall Effect \em.

\section{Perturbation theory}

We consider the light scattered by one dielectric sphere made
of a Faraday-active medium embedded in an isotropic medium with no
magneto-optical properties, using a perturbative approach to first order
in magnetic field. 

In this paper we set \( c_{0}=1 \). In a magnetic field, the refractive index
of the sphere is a tensor of rank two. 
It depends on the distance to the center
of the sphere \( r \), which is given a radius \( a \) via the Heaviside function \( \Theta (r-a), \)
that equals to 1 inside the sphere and 0 outside, \bfseries 

\begin{equation}
\varepsilon ({\mathbf{B}},{\mathbf{r}}) -{\mathbf{I}}  =  \left[ (\varepsilon _{0}-1)\, 
{\mathbf{I}} + \varepsilon_{F}\, \mathbf{\Phi} \right] \Theta (|{\mathbf{r}}|-a)\, .
\end{equation}

\mdseries In
this expression, \( \varepsilon _{0}=m^{2} \) is the value of the normal isotropic dielectric
constant of the sphere of relative index of refraction \( m \) ( which can
be complex ) and \( \varepsilon_{F}=2mV_{0}/\omega \) is a coupling parameter associated with the amplitude
of the Faraday rotation ( \( V_{0} \) being the Verdet constant and \( \omega  \) the frequency
). We introduce the antisymmetric hermitian tensor \( \Phi _{ij}=i\epsilon _{ijk}B_{k} \). 
Except for \( \varepsilon_{F} \), the Mie solution depends 
on the dimensionless size parameters \( x=ka \) with 
\( k=\omega/c_{0} \) and \( y=mx \). In this paper, we restrict ourselves
to non-absorbing media, so that \( m \) and \( \varepsilon_{F} \) are real-valued.

Noting that the Helmholtz equation is formally analogous to a Schr\"{o}dinger
equation with potential \( \mathbf{V}({\mathbf{r}},\omega )=\left[ {\mathbf{I}}-\varepsilon ({\mathbf{B}},{\mathbf{r}})\right] \, \omega ^{2} \) and energy \( \omega ^{2} \), the T-matrix is given by the
following Born series \cite{newton}:

\begin{equation}
\label{Born}
{\mathbf{T}}({\mathbf{B}},{\mathbf{r}},\omega )=\mathbf{V}({\mathbf{r}},\omega )+\mathbf{V}({\mathbf{r}},\omega )\cdot \mathbf{G}_{0}\cdot \mathbf{V}({\mathbf{r}},\omega )+...
\end{equation}
Here \( \mathbf{G}_{0}(\omega ,\mathbf{p})=
1/(\omega ^{2}{\mathbf{I}}-p^{2} \mathbf{\Delta} _{p}) \) 
is the free Helmholtz Green's function and \( (\Delta _{p})_{ij}=\delta _{ij}-p_{i}p_{j}/p^{2} \). We will call
\( {\mathbf{T}}^{0} \) the part of \( {\mathbf{T}} \) that is independent of the magnetic field and \( {\mathbf{T}}^{1} \) the
part of the T-matrix linear in \( {\mathbf{B}} \). It follows from Eq. (\ref{Born}) that:

\begin{equation}
\label{defpsi}
{\mathbf{T}}^{1}=(1-\varepsilon _{0})\varepsilon _{F}\Theta \, {\mathbf{T}}^{0}
\cdot \mathbf{\Phi}\cdot ({\mathbf{I}}+\mathbf{G}_{0}\cdot {\mathbf{T}}^{0}).
\end{equation}
We need to introduce the unperturbed eigenfunctions 
\( \Psi _{\sigma ,{\mathbf{k}}}^{\pm}\, ({\mathbf{r}} ) \) of the conventional
Mie problem. These eigenfunctions represent the electric field at the
point \( {\mathbf{r}} \) for an incident plane wave \( |\, \sigma ,{\mathbf{k}}> \)
 along the direction \( {\mathbf{k}} \) with an
helicity \( \sigma  \). This eigenfunction is "outgoing" for
\( \Psi _{\sigma ,{\mathbf{k}}}^{+} \) and "ingoing" for
 \( \Psi _{\sigma ,{\mathbf{k}}}^{-} \) according to the definition of the
  outgoing and
ingoing free Helmholtz Green's function: 
\[
|\, \Psi _{\sigma ,{\mathbf{k}}}^{\pm }>=({\mathbf{I}}+\mathbf{G}^{\pm }_{0}
\cdot {\mathbf{T}}^{0})\, |\, \sigma ,{\mathbf{k}}>.
\]
For our free Helmholtz Green's function, this implies \cite{newton}:
\begin{equation}
\Psi _{\sigma ,\mathbf{-k}}^{-*}( {\mathbf{r}} )=(-1)^{1+\sigma} \, 
\Psi _{-\sigma ,{\mathbf{k}}}^{+}( {\mathbf{r}} ).
\end{equation}

We denote by \( {\mathbf{k}} \) the incident direction and \( {\mathbf{k}}' \) the scattered direction.
With these notations, it is possible to obtain from Eq.(\ref{defpsi}),
 
\begin{equation}
\label{psi}
{\mathbf{T}}_{{\mathbf{k}}\sigma ,{\mathbf{k}}'\sigma '}^{1}=\varepsilon _{F}\, \omega ^{2}
<\Psi ^{-}_{\sigma ,{\mathbf{k}}}\, |\, \Theta \, \mathbf{\Phi} \, |\, \Psi _{\sigma ',{\mathbf{k}}'}^{+}>.
\end{equation}
This equation can also be obtained from standard first-order Rayleigh-Schr\"{o}dinger
perturbation theory \cite{landau}.

Two important symmetry relations are found to be satisfied by our T-matrix. The first
one is the parity and the second one is the reciprocity:
\begin{equation}
\label{parity}
{\mathbf{T}}_{-{\mathbf{k}}-\sigma ,-{\mathbf{k}}'-\sigma'}({\mathbf{B}})
={\mathbf{T}}_{{\mathbf{k}}\sigma ,{\mathbf{k}}'\sigma '}({\mathbf{B}})
\end{equation}
\begin{equation}
\label{reciprocity}
{\mathbf{T}}_{-{\mathbf{k}}'\sigma' ,-{\mathbf{k}}\sigma}(\mathbf{-B})
={\mathbf{T}}_{{\mathbf{k}}\sigma ,{\mathbf{k}}'\sigma '}({\mathbf{B}})
\end{equation}

\section{T matrix for Mie scattering}

In order to separate the radial and the angular contributions in Eq. (\ref{psi}),
we expand the Mie eigenfunction \( \Psi _{\sigma ,{\mathbf{k}}}^{+} \) 
in the basis of the vector spherical harmonics \cite{newton},
\begin{equation}
\label{def}
\Psi _{\sigma ,{\mathbf{k}}}^{+}\, ({\mathbf{r}})=\frac{2\pi }{\rho }\, i^{J+1}
{\mathbf{Y}}^{\lambda '}_{JM}\, (\hat {{\mathbf{r}}} )\, f^{J}_{\lambda '\lambda }(r)
\, {\mathbf{Y}}^{\lambda *}_{JM}(\hat {{\mathbf{k}}} )\cdot {\mathbf{\chi}}_{\sigma }'.
\end{equation}
In this definition, \( \rho =kmr \), the \( {\mathbf{\chi}} _{\sigma }' \) are the eigenvectors of the spin operator
in the circular basis associated with the direction \( {{\mathbf{k}}} \), and implicit
summation over the repeated indices \( J,M,\lambda  \) and \( \lambda ' \)
 has been assumed
 ( the indices \( \lambda  \) and \( \lambda ' \) for the components 
 of the field can take three values, 
one being longitudinal and two of them being perpendicular
to the direction of propagation).

Because of the presence of the function \( \Theta  \) in Eq. (\ref{psi}), we only
have to consider the field inside the sphere, whose main features
are contained in the radial function \( f^{J}_{\lambda '\lambda }(r) \) .
This matrix \( f^{J}_{\lambda '\lambda } \) is known in
terms of the transmission coefficients \( c_{J} \) and \( d_{J} \) 
of Ref. \cite{vanhulst}, and of the Ricatti-Bessel function
\( u_{J}(\rho ) \). We found the following expression:

\[
{\mathbf{f}}^{J}(\rho )=m\, \left( \begin{array}{ccc}
-iu_{J}'(\rho )c_{J} & 0 & 0\\
-iu_{J}'(\rho )c_{J}\, \sqrt{J(J+1)}/\rho  & 0 & 0\\
0 & 0 & u_{J}(\rho )d_{J}
\end{array}
\right) .\]

The three vectors \( {\mathbf{\chi}} _{\sigma} \) - similar to the \( {\mathbf{\chi}} _{\sigma }' \) 
but associated with the $z$-axis - are a convenient basis for 
this problem since they are the eigenvectors
of the operator \( {\mathbf{\Phi}} \) with eigenvalue \( -\sigma  \), provided we choose the $z$-axis
along \( {{\mathbf{B}}} \), which we will do in what follows. Eq. (\ref{psi}) is simplified by 
this choice and the angular integration leads eventually to:

\[
\int {\mathbf{Y}}^{\lambda _{1}*}_{J_{1}M_{1}}(\hat {{\mathbf{r}}} )\cdot {\mathbf{\Phi}}\cdot
 {\mathbf{Y}}^{\lambda _{2}}_{J_{2}M_{2}}(\hat {{\mathbf{r}}} )\, d\Omega _{r}=\delta _{J_{1}J_{2}\, }\delta _{M_{1}M_{2}\, }Q_{\lambda _{1}\lambda _{2}}(J_{1},M_{1}),\]
where \( {\mathbf{Q}} \) is the matrix

\begin{equation}
\label{matriceA}
{\mathbf{Q}}(J,M)=-MB\, \left( \begin{array}{ccc}
\frac{1}{J(J+1)} & \frac{1}{\sqrt{J(J+1)}} & 0\\
\frac{1}{\sqrt{J(J+1)}} & 0 & 0\\
0 & 0 & \frac{1}{J(J+1)}
\end{array}
\right) ,
\end{equation}
and \( B \) being the absolute value of the applied magnetic field. The linear
 dependence on the magnetic quantum number \( M \) can be expected
  for an effect like the Faraday rotation, affecting left and 
  right circularly polarized light in an opposite way similar 
  to Zeeman splitting.
The radial integration can be done using a method developed by Bott
et al. \cite{bott}. 

\begin{equation}
\label{T1}
{{\mathbf{T}}}_{{{\mathbf{k}}},{{\mathbf{k}}}'}^{1}=\frac{16\pi }{\omega }
\, W\, \sum _{J,M}\, (-M)\, \left[ \mathcal{C}_{J}\, 
{\mathbf{Y}}_{J,M}^{e}(\hat {{\mathbf{k}}} )\, {\mathbf{Y}}^{e*}_{J,M}(\hat {{\mathbf{k}}} ')
+\mathcal{D}_{J}\, {\mathbf{Y}}^{m}_{J,M}(\hat {{\mathbf{k}}} )\, {\mathbf{Y}}^{m*}_{J,M}(\hat 
{{\mathbf{k}}} ')\right] ,
\end{equation}
with the dimensionless parameter:

\[
W=V_{0}B\lambda ,\]
and the coefficients:

\begin{equation}
\label{C}
\mathcal{C}_{J}=-\frac{2\, c^{2*}_{J}|u_{J}|^{2}y^{3}}{J(J+1)\, (y^{2}-y^{*2})}\, 
\left( \frac{A^{*}_{J}}{y^{*}}-\frac{A_{J}}{y}\right) ,
\end{equation}

\begin{equation}
\label{D}
\mathcal{D}_{J}=-\frac{2\, d^{2*}_{J}|u_{J}|^{2}y^{3}}{J(J+1)\, (y^{2}-y^{*2})}\, \left( \frac{A^{*}_{J}}{y}-\frac{A_{J}}{y^{*}}\right) ,
\end{equation}
with \( A_{J}(y)=u_{J}'\, (y) / u_{J}\, (y) \) and \( B \) 
the amplitude of the magnetic field directed along the
unit vector \(  \hat {{\mathbf{B}}} \) . Absorption in the sphere is still allowed. We
will consider the limiting case of a perfect dielectric sphere with
no absorption ( \( \Im m(m)\rightarrow 0 \) ). Using l'Hospital's rule in Eqs. (\ref{C}) and (\ref{D}),
we obtain immediately for this case:

\begin{equation}
\label{C2}
\mathcal{C}_{J}=\frac{-c^{2*}_{J}u_{J}^{2}y}{J(J+1)}\, \left( \frac{A_{J}}{y}-\frac{J(J+1)}{y^{2}}+1+A^{2}_{J}\right) ,
\end{equation}

\begin{equation}
\label{D2}
\mathcal{D}_{J}=\frac{-d^{2*}_{J}u_{J}^{2}y}{J(J+1)}\, \left( -\frac{A_{J}}{y}-\frac{J(J+1)}{y^{2}}+1+A^{2}_{J}\right) .
\end{equation}

\subsection{T-matrix without magnetic field}

For future use, we need the on-shell T-matrix of the conventional Mie-problem
\cite{vanhulst}. It is given by a formula analogous to Eq. (\ref{T1}) where
\( \mathcal{C}_{J} \) and \( \mathcal{D}_{J} \) are replaced by the Mie coefficients \( a_{J} \) and \( b_{J} \), and with
\( M=1 \). Because of rotational invariance of the scatterer, it's clear that
the final result only depends on the scattering angle \( \theta  \) which is the
angle between \( {{\mathbf{k}}} \) and \( {{\mathbf{k}}}' \)
( see Fig. \ref{schema} ). Therefore, we get in the circular basis ( associated
with the indices \( \sigma  \) and \( \sigma ' \) ):

\begin{equation}
\label{t0}
T_{\sigma \sigma '}^{0}=\frac{2\pi}{i\omega }\sum _{J\geq 1}
\frac{2J+1}{J(J+1)}\, (a^{*}_{J}+\sigma \sigma 'b^{*}_{J})\, \left[ 
\pi _{J,1}(\cos \theta )+\sigma \sigma '\tau _{J,1}(\cos \theta )\right] .
\end{equation}
In this formula, the polynomials \( \pi _{J,M} \) and \( \tau _{J,M} \) 
are defined in terms
of the Legendre polynomials \( P^{M}_{J} \) by \cite{vanhulst}

\begin{equation}
\label{poly}
\pi _{J,M}(\cos \theta )=\frac{M}{\sin \theta }P^{M}_{J}(\cos \theta )\, ,\, 
\, \, \, \, \, \, \, \, \, \, \, \, \, \, \, \, 
\tau _{J,M}(\cos \theta )=\frac{d}{d\theta }P^{M}_{J}(\cos \theta ).
\end{equation}

\subsection{ \protect\( {{\mathbf{T}}}^{1}_{{{\mathbf{k}}} {{\mathbf{k}}}'} \protect \) 
when \protect\( \hat {{\mathbf{k}}} \neq \hat {\mathbf{k}} '\protect \)}

It remains to express the vector spherical harmonics in Eq. (\ref{T1}) in
terms of the natural angles of the problem in the presence of a magnetic
field. The latter breaks rotational invariance. Because 
\bfseries \( {\mathbf{T}}^{1} \) \mdseries is linear in \( \hat {\mathbf{B}}  \),
it can be constructed by considering
three special cases for the direction of \( \hat {\mathbf{B}}  \). 
If \( \hat {\mathbf{k}} \neq \hat {\mathbf{k}} ' \), we can decompose
the unit vector \( \hat {\mathbf{B}}  \) in the non-orthogonal but 
complete basis of \( \hat {\mathbf{k}} ,\, \hat {\mathbf{k}} ' \) and
\( \hat {\mathbf{g}} =\hat {\mathbf{k}} \times \hat {\mathbf{k}} '/|\hat 
{\mathbf{k}} \times \hat {\mathbf{k}} '| \), and this results in:

\begin{eqnarray}
\label{decomp} 
{\mathbf{T}}^{1}_{{\mathbf{k}} {\mathbf{k}}'}  = 
 \frac{(\hat {\mathbf{B}} \cdot \hat {\mathbf{k}} )(\hat {\mathbf{k}} 
 \cdot \hat {\mathbf{k}} ')-\hat {\mathbf{B}} \cdot \hat {\mathbf{k}} '}
 {(\hat {\mathbf{k}} \cdot \hat {\mathbf{k}} ')^{2}-1}
 {\mathbf{T}}_{\hat {\mathbf{B}} =\hat {\mathbf{k}} '}^{1} 
   +   \frac{(\hat {\mathbf{B}} \cdot \hat {\mathbf{k}} ')
   (\hat {\mathbf{k}} \cdot \hat {\mathbf{k}} ')-\hat {\mathbf{B}} \cdot 
   \hat {\mathbf{k}} }{(\hat {\mathbf{k}} \cdot \hat {\mathbf{k}} ')^{2}-1}
   {\mathbf{T}}_{\hat {\mathbf{B}} =\hat {\mathbf{k}} }^{1}
   +   (\hat {\mathbf{B}} \cdot {\mathbf{\hat g}} ) {\mathbf{T}}_{\hat {\mathbf{B}}
    =\mathbf{ \hat g} }^{1},  
\end{eqnarray}
with 
\begin{equation}
\label{tbpk}
T_{\sigma \sigma'}^{1} (\hat {\mathbf{B}} =\hat {\mathbf{k}} )=\frac{2W}
{\omega }\sum _{J\geq 1}\frac{2J+1}{J(J+1)}(-\sigma )\, 
(\mathcal{C}_{J}+\sigma \sigma '\mathcal{D}_{J})\, \left[ 
\pi _{J,1}(\cos \theta )+\sigma \sigma '\tau _{J,1}(\cos \theta )\right] ,
\end{equation}

\begin{equation}
\label{tbpkp}
T_{\sigma \sigma'}^{1} (\hat {\mathbf{B}} =\hat {\mathbf{k}} ')=\frac{2W}
{\omega }\sum _{J\geq 1}\frac{2J+1}{J(J+1)}(-\sigma ')\, 
(\mathcal{C}_{J}+\sigma \sigma '\mathcal{D}_{J})\, \left[ 
\pi _{J,1}(\cos \theta )+\sigma \sigma '\tau _{J,1}(\cos \theta )\right] ,
\end{equation}

\begin{equation}
\label{tt}
T_{\sigma \sigma'}^{1} (\hat {\mathbf{B}} =\hat {\mathbf{g}} )=\frac{4iW}
{\omega }\sum _{{J\geq 1\atop \, \, J\geq M>0}}\frac{2J+1}{J(J+1)}M\, \sin (M\theta )\, \frac{(J-M)!}{(J+M)!}(\sigma \sigma '\mathcal{C}_{J}+\mathcal{D}_{J})\, 
\left[ \pi_{J,M}^2(0)+\sigma \sigma '\tau_{J,M}^2(0)\right] .
\end{equation}

\subsection{Particular case for \protect\( {\mathbf{T}}^{1}\protect \) 
for forward scattering}

The treatment in (3.2) becomes degenerate when \( \hat {\mathbf{k}}  \) and \( \hat {\mathbf{k}} ' \) are parallel,
\em i.e. \em in forward scattering. 
In this case, \( \hat {\mathbf{B}}  \) can still be expressed
in a basis made of \( \hat {\mathbf{k}}  \) and of two vectors perpendicular to \( \hat {\mathbf{k}}  \). The contribution
of these last two vectors must have the same form as in Eq. (\ref{tt}) for \( \theta =0 \). Hence
there is no such contribution and we find

\begin{equation}
\label{tkk}
T_{{\mathbf{k}}\sigma ,{\mathbf{k}}\sigma '}^{1}=\delta _{\sigma \sigma '}\,
 (\hat {\mathbf{B}} \cdot \hat {\mathbf{k}} ) (-\sigma ) \, \frac{2W}
 {\omega }\sum _{J\geq 1}(2J+1)\, (\mathcal{C}_{J}+\mathcal{D}_{J}).
\end{equation}
In Fig. \ref{f:tmatrix} we plotted real and imaginary part of this expression
in units of \( W \) as a function of the size parameter \( x \) for 
\( \sigma=-1 \) and \( \hat {\mathbf{B}}=\hat {\mathbf{k}} \).
The forward-scattering amplitude has an important application in
inhomogeneous media, namely as the complex average dielectric constant.

\subsection{Optical Theorem}

We will check our formula on energy conservation as expressed by the
Optical Theorem \cite{newton},

\begin{equation}
\label{opticalth}
-\frac{\Im m(T_{{\mathbf{k}}\sigma ,{\mathbf{k}}\sigma })}{\omega }=\sum _{\sigma '}\, \int d\Omega _{{\mathbf{k}}'}\, 
\frac{|T_{{\mathbf{k}}\sigma ,{\mathbf{k}}'\sigma '}|^{2}}
      {(4\pi)^2}.
\end{equation}
To first order in magnetic field, the r.h.s of this equation equals 
\[
\frac{1}{8\pi^2}
\sum _{\sigma '}\int d\Omega _{{\mathbf{k}}'}\, \Re e(T^{0}_{{\mathbf{k}}\sigma ,{\mathbf{k}}'\sigma '}\, T_{{\mathbf{k}}\sigma ,{\mathbf{k}}'\sigma '}^{1*}).\]
If we assume that \( \hat {\mathbf{B}} \parallel \hat {\mathbf{k}}  \), we can compute this using Eqs. (\ref{tbpk}), (\ref{t0}), and
the following orthogonality relations for the polynomials \( \pi _{J,1} \) and \( \tau _{J,1} \)
( which we denote as \( \pi _{J} \) and \( \tau _{J} \) ) \cite{newton} :

\[
\int d(\cos \theta )\, [\pi _{J}(\cos \theta )\tau _{K}(\cos \theta )+\tau _{J}(\cos \theta )\pi _{K}(\cos \theta )]=0,\]

\begin{equation}
\label{orthogonalite}
\int d(\cos \theta )\, [\pi _{J}(\cos \theta )\pi _{K}(\cos \theta )+
\tau _{J}(\cos \theta )\tau _{K}(\cos \theta )]=\frac{2J^{2}(J+1)^{2}}{2J+1}\,
 \delta _{JK}.
\end{equation}

The l.h.s of Eq. (\ref{opticalth}), is obtained from Eq. (\ref{tkk}). The Optical Theorem
provides us a relation between Mie coefficients, which we can
actually prove from their definitions:

\begin{equation}
\label{newrel}
\Re e(a_{J}^{*}c^{2}_{J}\, \frac{2}{i})=\Im m(c^{2}_{J}),
\end{equation}

and

\begin{equation}
\label{newrel2}
\Re e(b_{J}^{*}d^{2}_{J}\, \frac{2}{i})=\Im m(d^{2}_{J}).
\end{equation}

\section{Magneto-transverse Scattering}

From the knowledge of the matrix \( {\mathbf{T}}^{1} \), we can compute how the magnetic
field affects the differential scattering cross section summed over
polarizations as a function of the scattering angle. Only the diagonal
part of this matrix in a basis of linear polarization will affect
the scattering cross section since we consider only terms to first order in
magnetic field. Therefore only the third contribution of
Eq. (\ref{decomp}) plays a role, which means that the effect
will be maximum when \( \hat {\mathbf{B}} \parallel \hat {\mathbf{k}} \times
 \hat {\mathbf{k}} ' \) ( the typical Hall geometry ).
Indeed symmetry implies that the magneto-scattering cross section 
should be the product of \( \det(\hat {\mathbf{B}},\hat {\mathbf{k}},\hat {\mathbf{k}}') \)
and of some function that entirely depends on the scattering angle \( \theta \).

We choose to normalize this magneto scattering cross section  
by the total scattering cross section in the absence of magnetic field,

\begin{equation}
\label{paral}
\frac{2\, \sum _{\sigma \sigma '}\Re e(T_{\sigma \sigma '}^{0}
 T_{\sigma \sigma '}^{1*})}
{\int ^{2\pi}_{0} d\phi \int ^{\pi }_{0}d\cos \theta \, 
\sum _{\sigma \sigma '}|T^{0}_{\sigma \sigma '}|^{2}}
=-\sin \phi F(\theta )
\end{equation}

The so called \em Photonic Hall effect \em (P.H.E) is a manifestation
of a magnetically induced transverse current in the light transport
which has similarities to the Hall effect known for the transport
of electrons. In an experiment on the \em \em P.H.E, one measures
the difference in scattered light from two opposite directions both
perpendicular to the incident direction of light and to the applied
magnetic field \cite{nature}. The P.H.E is a manifestation of
the anisotropy of light scattering due to a magnetic field in
the regime of multiple light scattering.

Although experiments dealt with multiple scattering so far, it is
interesting to see if a net magneto-transverse scattering
 persists for only one single scatterer. In Figs. \ref{fig2},\ref{fig1},
 the magnetic field is perpendicular to the plane of the figure, 
 the incident light is along the $x$-axis. A typical measurement of the
 magneto-transverse scattered light is therefore 
 associated with the projection of the
curve onto the $y$-axis, which we define as the magneto-transverse direction.

\section{Magneto-transverse scattering as a function of the size parameter}

Quantitatively, the transverse light current difference is associated
with a summation of the magneto scattering cross section over outgoing
wavevectors and normalized to the total transverse light current. A
schematic view of the geometry is displayed on Fig. \ref{schema}.
In our notation this is

\begin{equation}
\label{hall}
\eta \equiv \frac{ I_{up}(B)-I_{down}(B) }{I_{up}(B=0)+I_{down}(B=0)}=\frac
{2\int ^{\pi }_{0}d\phi
\int ^{\pi }_{0}d\cos \theta \, \sin \theta \sin \phi 
\, \sum _{\sigma \sigma '}\Re e(T_{\sigma \sigma '}^{0} T_{\sigma \sigma '}^{1*})}
{\int ^{\pi }_{0}d\phi
\int ^{\pi }_{0}d\cos \theta \, \sin \theta \sin \phi \, 
\sum _{\sigma \sigma '}|T^{0}_{\sigma \sigma '}|^{2}}.
\end{equation}
The factor \( \sin\theta \sin \phi \) represents a projection onto the
magneto transverse direction 
\( \hat {\mathbf{B}} \times \hat {\mathbf{k}} \) which is necessary 
since we are interested in the magneto-transverse light flux.
In Fig. \ref{fig3} we plotted this contribution as a function of the size
parameter \( x \) for an index of refraction
\( m=1.0946 \) ( the value in Ref. \cite{nature} ). Note the change of sign
beyond \( x=1.7 \), for which we do not have any simple explanation so far.
In the range of small size parameter, \( \eta \) exhibits an
\( x^5 \) dependence.

\subsection{Rayleigh scatterers}

For Rayleigh scatterers, the formulas (\ref{decomp}) to 
(\ref{tkk}) simplify dramatically because
one only needs to consider the first partial wave of \( J=1 \) and the first
terms in a development in powers of \( y \) ( since \( y\ll 1 \) ). From Eqs. (\ref{C2}) and
(\ref{D2}), we find :
\( \mathcal{C}_{1}=-2y^{3}/m^{2}(2+m^{2})^2 \) and
 \( \mathcal{D}_{1}=-y^{5}/45m^{4} \) so that we can keep only \( \mathcal{C}_{1} \) and drop \( \mathcal{D}_{1} \) as a first approximation.
Adding all contributions of Eqs. (\ref{decomp}) and (\ref{t0}), and changing from
circular basis to linear basis of polarization we find

\begin{equation}
\label{ttotal}
{\mathbf{T}}_{{\mathbf{k}},{\mathbf{k}}'}=\left( 
\begin{array}{cc}
t_0 \hat {\mathbf{k}} \cdot \hat {\mathbf{k}} '
+i t_1 \hat {\mathbf{B}} \cdot (\hat {\mathbf{k}} \times \hat {\mathbf{k}} ') 
& i t_1 \hat {\mathbf{B}} \cdot \hat {\mathbf{k}} \\
-i t_1 \hat {\mathbf{B}} \cdot \hat {\mathbf{k}} ' & 
t_0
\end{array}
\right) ,
\end{equation}
with \( t_0=-6i\pi a^{*}_{1}/ \omega \) the conventional Rayleigh
T-matrix and  \( t_1=6\mathcal{C}_{1}W / \omega \).
This form agrees with the Rayleigh point-like scatterer 
model discussed in Ref. \cite{bart}. 
We note that Eqs. (\ref{tbpk}) and (\ref{tbpkp}) give off-diagonal contributions in Eq.
(\ref{ttotal}) whereas Eq. (\ref{tt}) gives a diagonal contribution. This is a general
feature that persists also beyond the regime of 
Rayleigh scatterers.

For a Rayleigh scatterer the magneto cross section of 
Fig. \ref{fig2} exhibits symmetry, since
 the positive and negative lobes of the curve are of the same size
but of opposite sign. Hence no net magneto-transverse scattering
 exists for one Rayleigh scatterer.
 In fact Eq. (\ref{ttotal}) provides the
following expression for \( F(\theta) \):
\begin{equation}
\label{Frayleigh}
F_{ \mbox{Rayleigh} }(\theta)=\frac{3mx^3}{4\pi^2 (m^2+2)^2 } 
\, \cos\theta \sin\theta.
\end{equation}

 As the size of the sphere gets bigger, the magneto
corrections become asymmetric, as seen in Figure \ref{fig1}. When the size
is further increased, new lobes start to appear in the magnetic cross section
corresponding to higher spherical harmonics. These lobes do seem to
have a net magneto-transverse scattering.

One single Rayleigh scatterer does not induce a net magneto-transverse flux.
It is instructive to consider the next simplest case, namely two 
Rayleigh scatterers positioned at ${\mathbf{r}}_1$ and ${\mathbf{r}}_2$.
If their mutual  separation well exceeds the wavelength, it
is easy to show that the collective cross-section simply equals
the one-particle cross-section multiplied by an interference factor
$S( {\mathbf{k,k'}}) = \left|\exp(i({\mathbf{k}}-\mathbf{k'}) \cdot 
{\mathbf{r}}_1) + \exp(i({\mathbf{k}}-\mathbf{k'}) \cdot {\mathbf{r}}_2)
\right|^2$. This interference 
factor changes the angular profile of the scattering
cross-section and makes sure that a net magneto-transverse flux remains.
The estimate for two Rayleigh particles with an incident wave vector
along the inter-particle axis,
is found to be:
\begin{equation}
\eta \sim \frac{V_0 B}{k} \, x^3\, 
\left(\frac{\sin(kr_{12})}{kr_{12}}\right)^2 
\hspace{.5in} \mbox{if} \, \, \, \, kr_{12} \gg 1
\label{two}
\end{equation} 
This simple model suggests that the ``Photonic Hall Effect" is
in fact a phenomenon generated by interference of different
light paths.
In Fig.~\ref{anja} we show how differential cross-sections of two
particles change in a magnetic field. More scattering is now directed
into the forward direction and as a result the cancellation of
the net magneto-transverse flux in Fig.~\ref{fig2} is removed.   
One Mie sphere mimics qualitatively this simple model and should  
on the basis of the  principle outlined above
  exhibit a magneto-transverse current. The model also
suggests that the regime of Rayleigh-Gans scattering \cite{vanhulst} - the Born
approximation for one sphere but contrary to Rayleigh scattering 
still allowing interferences of different scattering events - 
should exhibit a  net magneto-transverse flux. Indeed, explicit 
calculations in this regime confirm this statement, with
$\eta \sim x^5$, independent of the
index of refraction $m$ of the sphere.

\subsection{Geometrical Optics}

In the regime of large size parameters, the Mie solution can be obtained from
Ray Optics. Apart from Fraunhofer diffraction
( persisting for any finite geometry ), the Mie solution
for a ray with central impact should be equivalent to the one for
a slab geometry. In our magnetic-optic approach, this means
to study Faraday rotation in a Fabry-Perot cavity. This model
is of special interest because the Fabry-Perot cavity is known to
enhance the Faraday rotation \cite{fp,fp2}, \em i.e \em
the rotation is additive in the total traversed path length
as opposed to the case of rotary power.

A ray with central impact is characterized by \( J=1 \) in 
Mie theory. We assume that \( x\gg 1 \) and \( y\gg 1 \) 
which allows some simplification in the expression of 
the Mie coefficients. We find 
for the Mie coefficients \( c_{1} \) and \( d_{1} \)
the following behavior:

\[
c_{1}=\frac{2\, e^{i(x-y)}}{(m+1)\, (1+re^{-2iy})},
\]

\begin{equation}
\label{c3}
d_{1}=\frac{2\, e^{i(x-y)}}{(m+1)\, (1-re^{-2iy})}.
\end{equation}

In this formula, \( r=(m-1)/(m+1) \) is the complex Fresnel reflection coefficient. Putting
this expression into Eqs. (\ref{C2}), (\ref{D2}) and (\ref{tkk}), we can compute the exact
behavior of the T-matrix in the forward direction. We note 
\( {\mathbf{T}}_{scatt}^{0} \) the part of \( {\mathbf{T}}^{0} \) 
due to scattering only, which is obtained from Eq. (\ref{t0}) by replacing
 the Mie coefficients \( a_{J} \) and \( b_{J} \) by \( a_{J}-\frac{1}{2} \) 
 and \( b_{J}-\frac{1}{2} \), since the terms \( \frac{1}{2} \) are associated
  with the Fraunhoffer diffraction, 
  which does not exist for the slab geometry.

Since for a small perturbation, we have

\begin{equation}
\label{approx}
{\mathbf{T}}={\mathbf{T}}_{scatt}^{0}+{\mathbf{T}}^{1} \simeq
{\mathbf{T}}^{0}_{scatt} e^{ {\mathbf{T}}^{1}/
 {\mathbf{T}}_{scatt}^{0}},
\end{equation}
we see that the change in phase \( \delta \phi  \) due to the magnetic field
is in fact related to the imaginary part of 
\( {\mathbf{T}}^{1}/{\mathbf{T}}_{scatt}^{0}. \) 
This change in phase can be
interpreted as the Faraday effect.

From Eq. (\ref{c3}), we find in the basis of circular polarization:

\begin{equation}
\label{defphi}
\Im m \left( \frac{T^{1}}{T_{scatt}^{0}} \right) _{\sigma \sigma'}=
\delta \phi \, (-\sigma)\hat {\mathbf{B}} \cdot \hat {\mathbf{k}} \,
\delta_{\sigma \sigma'},
\end{equation}
with
\begin{equation}
\label{fp}
\delta \phi =2aV_{0}B \frac{1+R}{1-R}\frac{1}
{[1+\mathcal{M}\sin (2y)^{2}]}.
\end{equation}
with \( \mathcal{M}=4R/(1-R)^{2} \) and the reflectivity \( R=r^{2}. \) 
We note that the quantity \( (-\sigma)\hat 
{\mathbf{B}} \cdot \hat {\mathbf{k}} \) is conserved for a given
ray, which generates the accumulation of the Faraday rotation.
The function \( \delta \phi  \) tends to \( 2aV_{0}B \) as \( R\rightarrow 0 \), 
since it represents the normal Faraday rotation in an isotropic medium of
length \( 2a \), as it should be for our geometry. When \( R \) is large, two new
factors come into play: \( (1+R)/(1-R) \) which is the maximum gain factor of the
Faraday rotation due the multiple interference in the Fabry-Perot
cavity and 
\[
\mathcal{A}(y)=\frac{1}{1+\mathcal{M}\sin (2y)^{2}}\]
which is an Airy function of width \( 4/ \sqrt{\mathcal{M}}. \) 
The finesse of the cavity is then 
\( \mathcal{F}=\pi \sqrt{\mathcal{M}}/2. \)
 At resonance, the Faraday rotation is maximally amplified
- assuming no losses - relative to single-path Faraday rotation
\cite{fp,fp2}. We stress that one needs
\( \delta \phi \ll 1 \), in order for Eq. (\ref{approx}) to apply. 
The Faraday rotation has the effect of splitting each 
transmission peak in the Fabry-Perot cavity into two
peaks of smaller amplitude both associated with a different state
of helicity.

The amplification of the Faraday rotation is a consequence
of the amplified path length of the light. In other words,
the Faraday rotation measures the time of interaction of the
 light with the magnetic field \cite{gasparian}. This time is
  found to be the \em dwell time \em \( \tau \) of the light 
  in the cavity for this 1D problem . 
The change in phase follows the simple relation:

\begin{equation}
\label{dwell}
\delta \phi =V_{0}B\frac{\tau c_{0}}{m}
\end{equation}
where \( c_{0}/m \) is recognized as the speed of light in the sphere.

The \em dwell time \em of the light in the cavity varies
between a maximum value of

\[
\tau ^{max}_{dwell}=(1+m^{2}) a /c_0 ,\]
 and a minimum value of

\[
\tau ^{min}_{dwell}=4m^{2}/(1+m^{2})\, a /c_0 .\]
These typical oscillations are visible in the plot of the
change of phase \( \delta \phi \) of Fig. \ref{figfp}.

\section{Summary and Outlook}

In this paper we addressed the Faraday effect inside a dielectric 
sphere. We have shown that this theory 
is consistent with former results concerning the predictions of the
light scattered by Rayleigh scatterers in a magnetic field. It is
possible to get from this perturbative theory quantitative predictions
concerning the \em Photonic Hall Effect \em for one single Mie sphere,
such as the scattering cross section, the dependence on the size parameter
or on the index of refraction. 

We will start experiments addressing single Mie scattering in a magnetic
field. A second challenge is to implement our Mie solution into a multiple
scattering theory.

\acknowledgments

We wish to thank Y. Castin for very stimulating and valuable discussions.


\newpage


\begin{figure}
\caption{Schematic view of the magneto-scattering geometry. Generally,
 $\theta$ denotes the angle between incident and outgoing wave vectors; $\phi$
 is the azymuthal angle in the plane of the magnetic field and
 the $y$-axis. The latter is by construction the magneto-transverse direction
 defined as the direction perpendicular to both
 magnetic field and incident wave vector. The angle $\alpha$ 
 coincides with the angle
 $\theta$ in the special but relevant case that the incident vector
 is normal to the magnetic field.  \label{schema}  }
\end{figure}

\begin{figure}
\caption{The solid line is the real part and the dashed line the imaginary
part of the magneto forward scattering matrix 
\protect\( {\mathbf{T}}_{\mathbf{kk}}^{1}\protect \) in the circular basis
of polarization, plotted versus the size parameter
\protect\( x\protect \) for an index of refraction 
of \protect\( m=1.33\protect \) in units of \protect\( W\protect=V_0 B \lambda \). 
\label{f:tmatrix} }
\end{figure}

\begin{figure}
\caption{Magneto-transverse scattering cross section \protect\( F(\theta )\protect \) 
for a Rayleigh scatterer with index of refraction 
\( m=1.1 \) and size parameter \( x=0.1 \). The solid
line is a positive correction and the points denote a negative correction.
The curve has been normalized by the parameter \protect\( W\protect \).
 \label{fig2} 
No net magneto-transverse scattering is expected in this case
 because the projection onto the $y$-axis of these corrections cancel.}
\end{figure}

\begin{figure}
\caption{Magneto-transverse scattering cross section \protect\( F(\theta )\protect \) 
for a Mie scatterer of size parameter \( x=5 \)\label{fig1} and 
of index of refraction \( m=1.1 \). The curve has been normalized 
by the parameter \protect\( W\protect \). The solid line
is for positive correction and the points denote a negative correction.
In this case a net magneto-transverse scattering is expected because
the projection onto the $y$-axis of these corrections do not cancel.}
\end{figure}

\begin{figure}
\caption{Normalized magneto-transverse light current \protect\( \eta \protect \) as a function of the
size parameter \protect\( x\protect \) for an index of refraction of \protect\( m=1.0946\protect \). The curve is displayed
in units of \protect\( W \). \label{fig3}}
\end{figure}

\begin{figure}
\caption{Magneto-cross section for two Rayleigh scatterers each of 
size parameter \( ka=0.1 \) and separated by a distance corresponding
to a size parameter of \( kr_{12}=5 \). In this case the enhanced forward
 scattering leads also to a net magneto-transverse current along
  the vertical axis.}
\label{anja}
\end{figure}

\begin{figure}
\caption{Magnetically induced change of phase \( \delta \phi \) 
- similar to Fabry-Perot modes of a cavity - as a function of size 
parameter \protect\( x\protect \) for the partial
wave of \protect\( J=1\protect \) - ``central impact'' -. 
The curve is 
for \protect\( m=10\protect \),  and has been normalized by the
value \( 2aV_0 B \).\label{figfp}}
\end{figure}

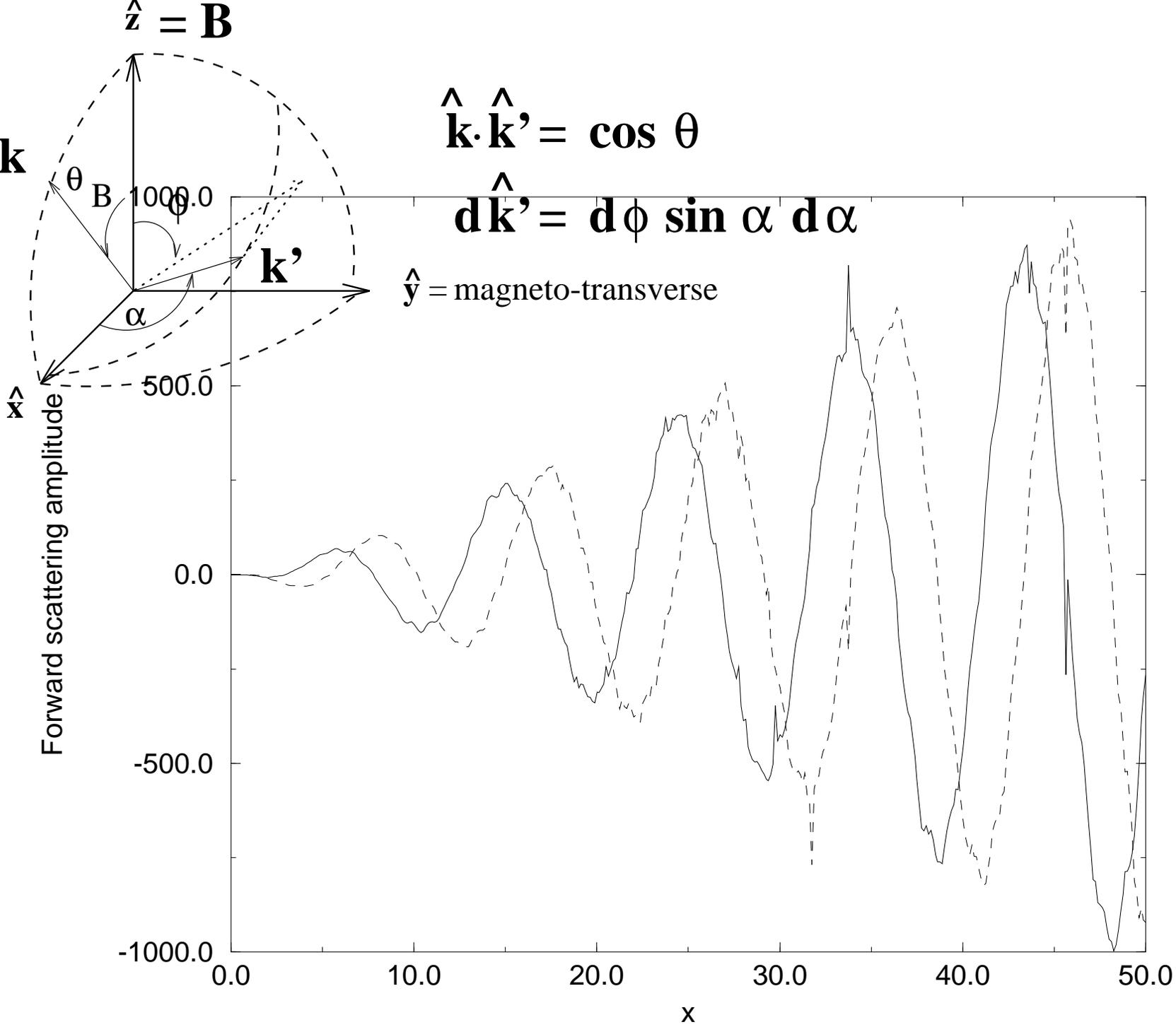
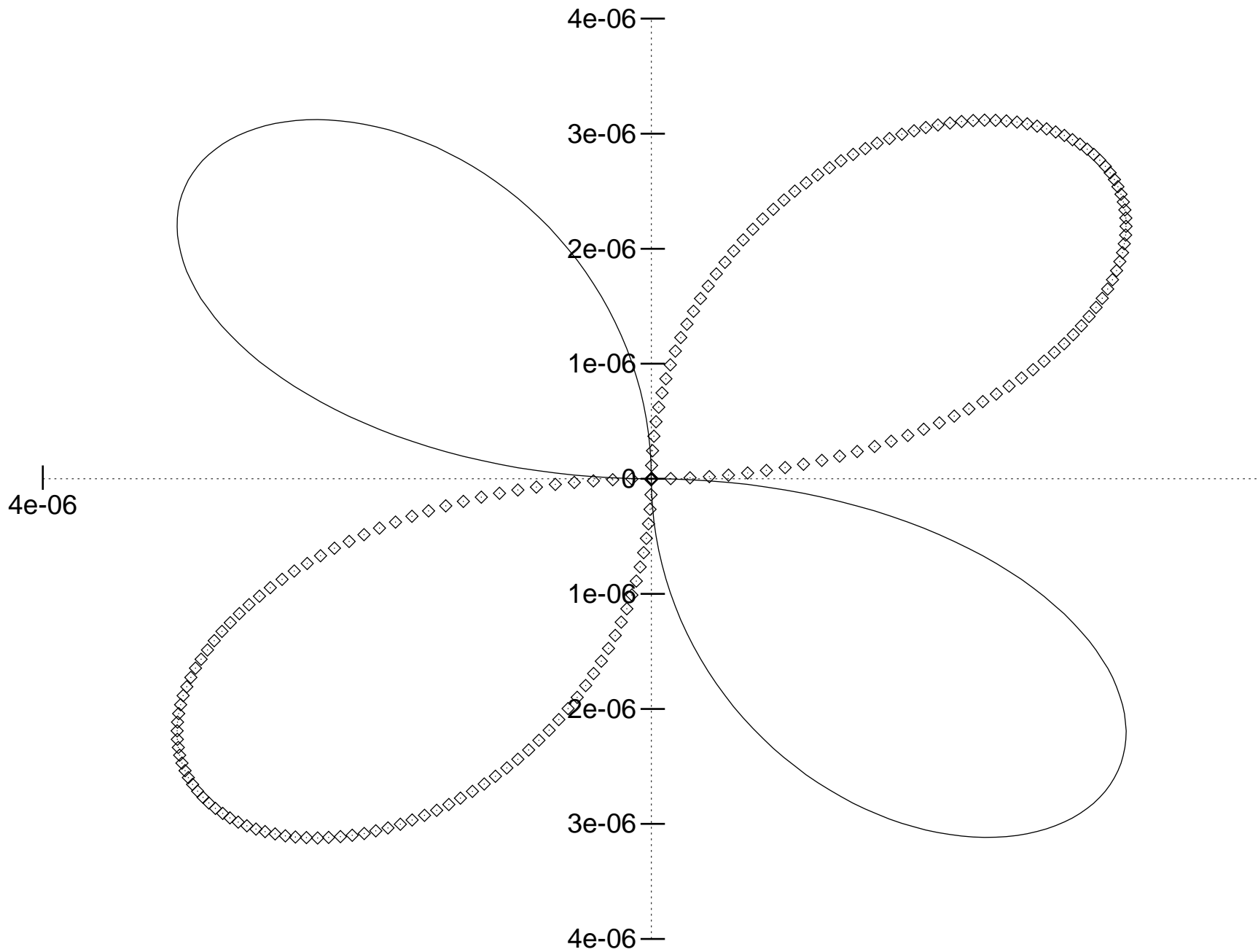
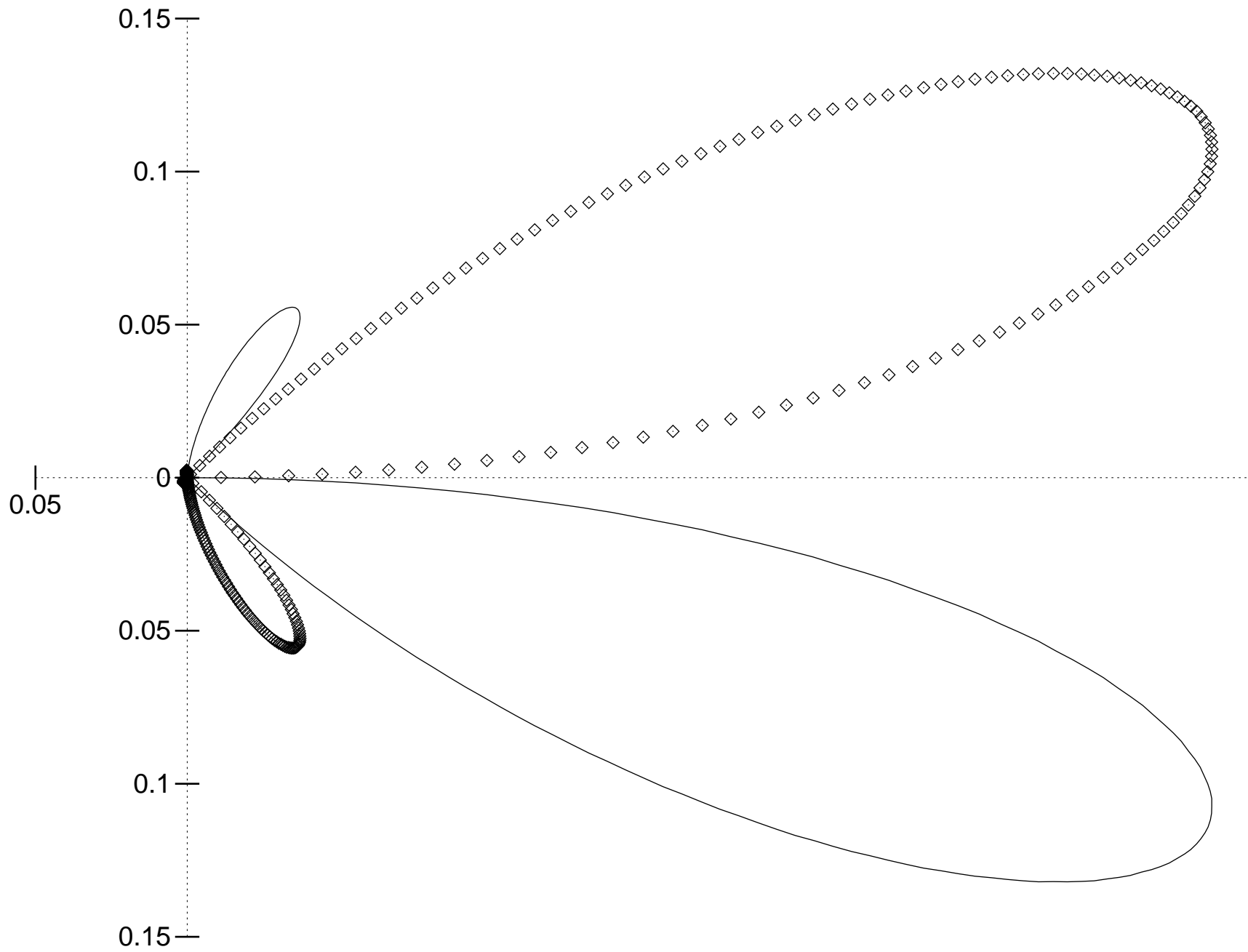
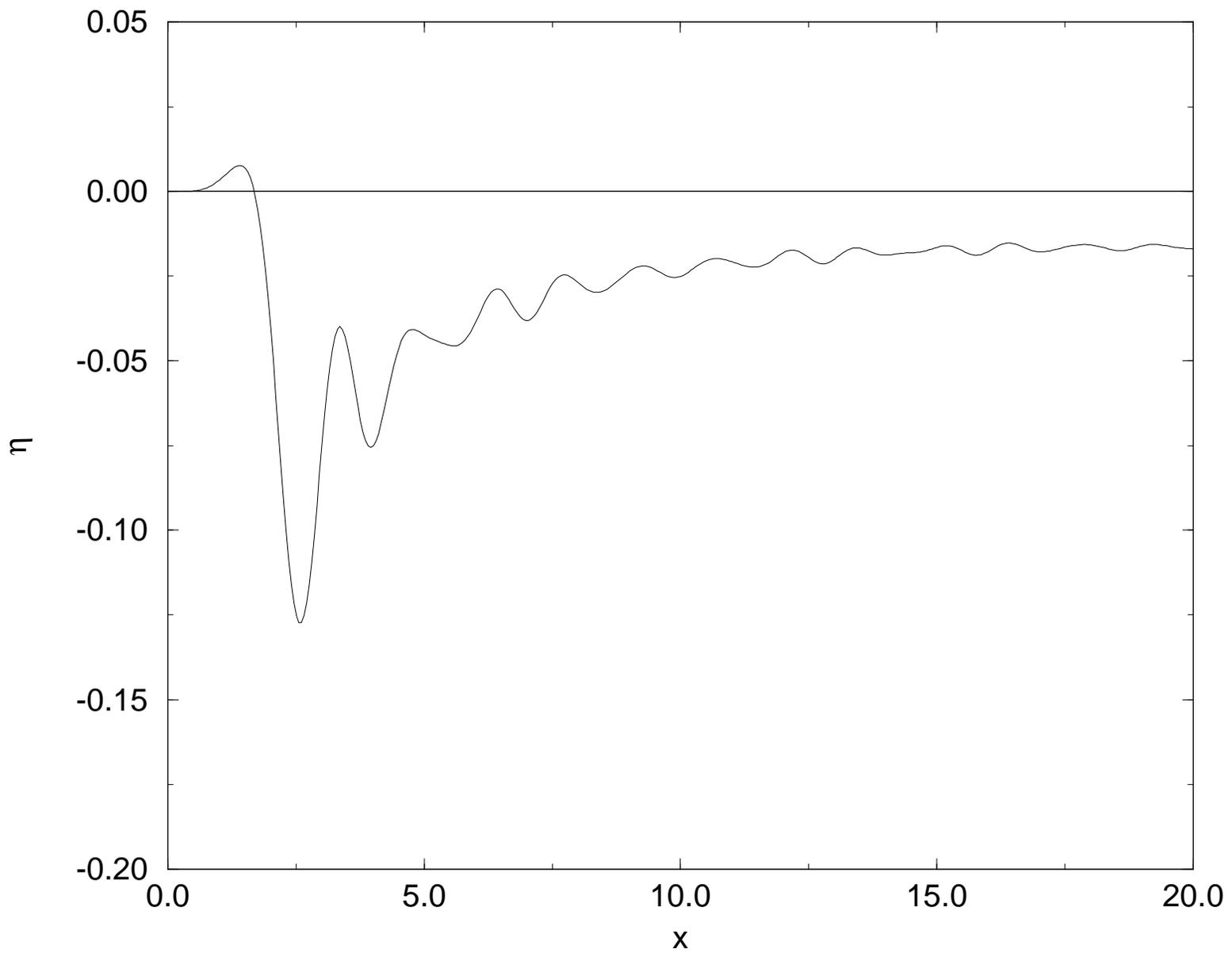
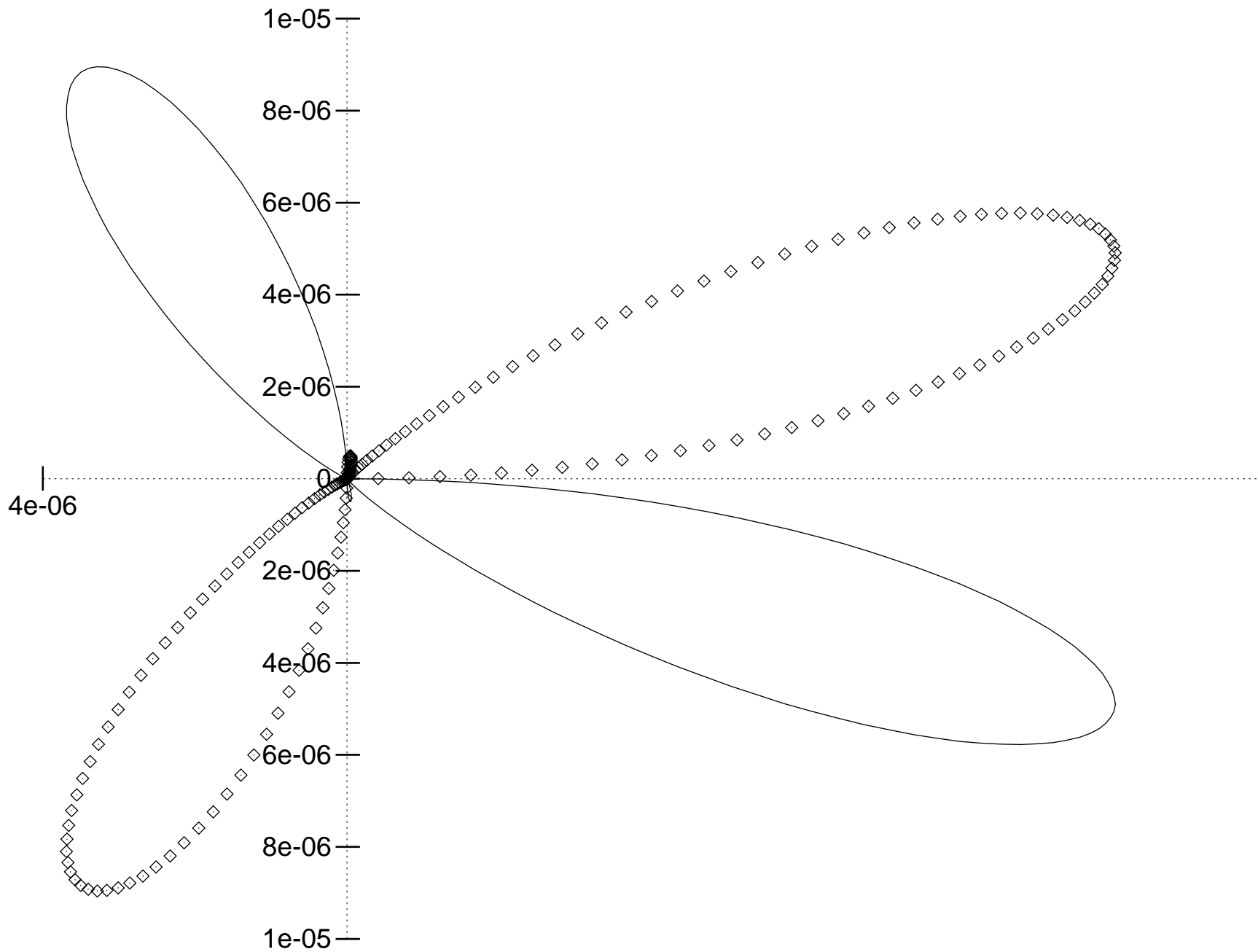
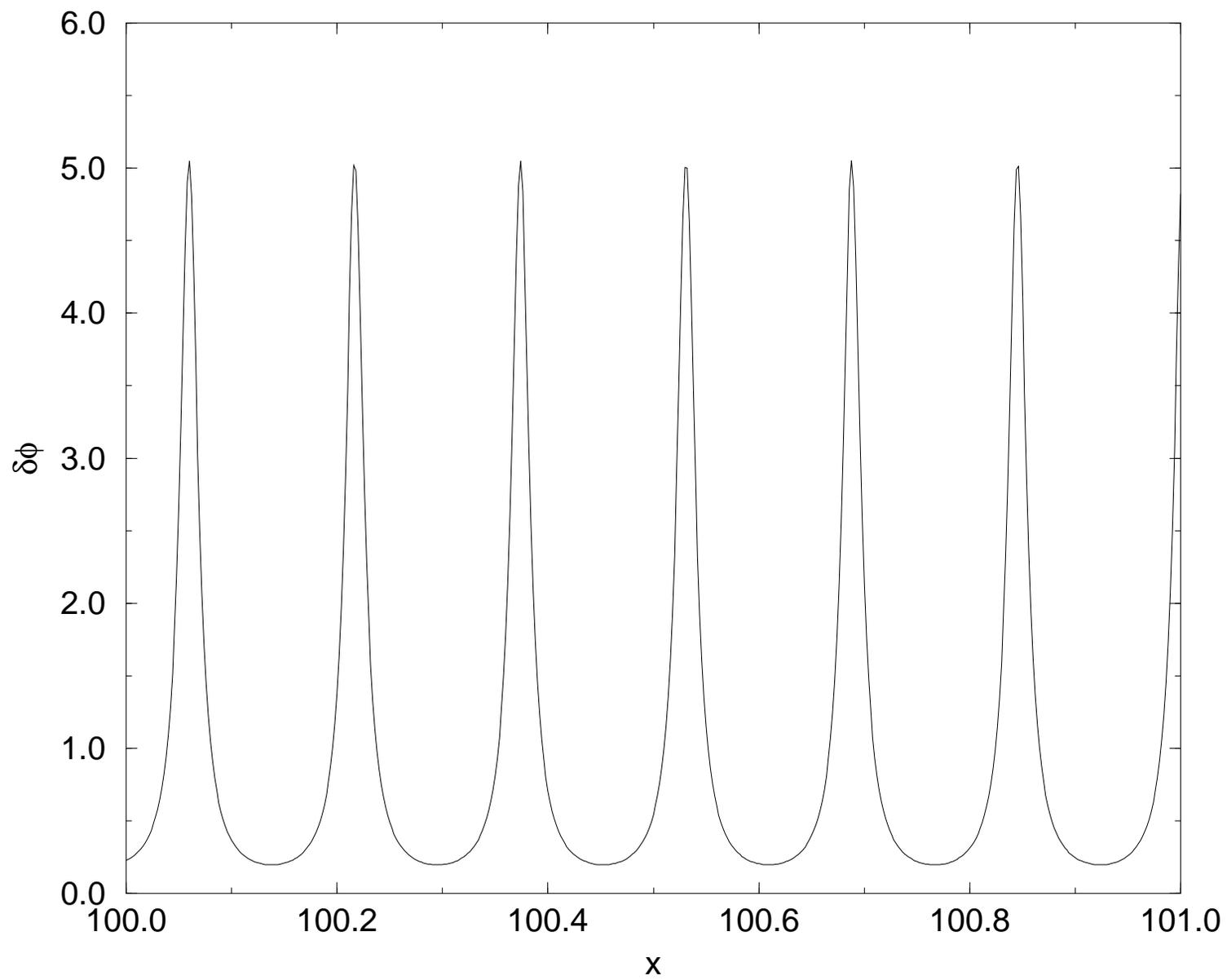
\end{document}